\documentclass[12pt,a4paper]{article}
\usepackage{latexsym}
\makeatletter
\@addtoreset{equation}{section}
\makeatother

\def\unit{\hbox to 3.3pt{\hskip1.3pt \vrule height 7pt width .4pt \hskip.7pt
\vrule height 7.85pt width .4pt \kern-2.4pt
\hrulefill \kern-3pt
\raise 4pt\hbox{\char'40}}}

\def\x{\times}

\def \alfa {(2 \pi \alpha')}
\def \ikC {({i}_{\hat k} {\hat C})}
\def \iktildeC {({i}_{\hat k} {\hat {\tilde C}})}
\def \ikchi {({i}_{\hat k} {\hat \chi})}
\def \iktildeN {({i}_{\hat k} {\hat {\tilde N}})}
\def \iktildechi {({i}_{\hat k} {\hat {\tilde \chi}})}

\begin{document}

\begin{flushright}
\footnotesize
UG-5/98\\
QMW-PH-98-07 \\
February, $1998$
\normalsize
\end{flushright}

\begin{center}


\vspace{.6cm}
{\LARGE {\bf The Massive Kaluza-Klein Monopole}}

\vspace{.9cm}

{
{\bf Eric Bergshoeff}
\footnote{E-mail address: {\tt E.Bergshoeff@phys.rug.nl}},
{\bf Eduardo Eyras}
\footnote{E-mail address: {\tt E.A.Eyras@phys.rug.nl}}\\
{\it Institute for Theoretical Physics\\
University of Groningen\\
Nijenborgh 4, 9747 AG Groningen, The Netherlands}
}

\vspace{.5cm}

{
{\bf Yolanda Lozano}
\footnote{E-mail address: {\tt Y.Lozano@qmw.ac.uk}}\\
{\it Physics Department\\
Queen Mary \& Westfield College\\
Mile End Road, London E1 4NS, U.K.}
}

\vspace{.2cm}


\vspace{.2cm}

\vspace{.8cm}


{\bf Abstract}

\end{center}

\begin{quotation}

\small

We construct the (bosonic) effective worldvolume action of an M-theory
Ka\-lu\-za--Klein monopole in a background given by the bosonic sector of
eleven--dimensional massive supergravity, i.e.~a ``massive Ka\-lu\-za--Klein
mo\-no\-pole''. As a consistency check we show that the direct dimensional
reduction along the isometry direction of the Taub--NUT space leads to
the massive D-6-brane. We furthermore perform a double dimensional
reduction in the massless case and obtain the effective 
worldvolume action of a Type IIA Kaluza--Klein monopole.

\end{quotation}

\vspace{1cm}

\newpage

\pagestyle{plain}


\section*{Introduction}

It is well-known that a classification of worldvolume (scalar, 
vector and tensor) multiplets leads to the brane scans for 
p-branes, D-branes and M-branes. On the other hand, requiring that 
branes in different dimensions are related to each other via
(direct or double) dimensional reduction leads one to include
gravitational waves and Kaluza--Klein (KK) monopoles \cite{Sorkin-Gross-Perry}
as well. These objects are missing
in the standard brane scans which are based upon the so-called
Bose--Fermi matching conditions. 
For the gra\-vi\-ta\-tio\-nal waves the reason is simply 
that their effective action is given by a one-dimensional field theory 
\cite{BJO} and for such low--dimensional systems the Bose--Fermi
matching conditions do not apply.

It turns out that the worldvolume action of a $d$--dimensional KK--monopole 
corresponds to a 
$(d-5)$--dimensional field theory \cite{Hull} and for such case the
Bose--Fermi matching conditions do apply. A natural question to ask
is then why is the KK--monopole missing in the standard brane scans?
As discussed in \cite{Hull}, a special feature of the KK--monopole
is that one of its four transverse directions corresponds to the isometry
direction in the Taub-NUT space of the monopole\footnote{For $d=11$,
a direct dimensional reduction over this isometry direction leads to the
D-6-brane. This does  not imply that the M--theory 
KK--monopole is equivalent to the ten--dimensional D-6-brane.
The difference is that the D-6-brane is a {\it singular} object moving in ten
dimensions whereas the M--theory KK--monopole is a {\it nonsingular} object
defined in eleven dimensions (of which one is compact). We thank Paul 
Townsend for pointing this out to us.}. 
Since the monopole cannot
move in this direction one should not associate a physical worldvolume scalar
to it. On the other hand, given the fact that a monopole is a
$(d-5)$--brane one cannot use the worldvolume diffeomorphisms
to gauge away this unphysical scalar. In order to write down a 
Lorentz--covariant action one therefore needs a new mechanism
to eliminate unphysical worldvolume scalars. We proposed in \cite{BJO} that the
effective action of the Heterotic KK--monopole corresponds to a 
{\it gauged sigma model} \cite{Hull-Spence}
with a gauging in the isometry direction of the Taub--NUT space.
The effect of this gauging is to eliminate the unwanted scalar.

The purpose of this letter is to extend the results of \cite{BJO}
(for other partial results on the KK--monopole see \cite{Imamura}) to
the Type IIA and M--theory KK--monopoles. For these cases the additional
feature arises that the KK--monopole may move in a background with a
non--zero cosmological constant proportional to $m^2$ with $m$ a mass
parameter. Branes moving in such a background are called ``massive branes''.
Their properties have been investigated in \cite{BLO} (for earlier 
results on massive branes, see \cite{YT}). It turns out that 
eleve--dimensional massive
branes are described by a {\it gaug\-ed sig\-ma model} with the gauge coupling
constant proportional to $m$. This letter extends the work of
\cite{BLO} to the case of the massive M-theory KK--monopole
\footnote{For some recent work on  KK--monopoles, see
e.g. \cite{some}.}.

At this point an obvious question arises. On the one hand, it follows
from \cite{BJO} that the massless M--theory KK--monopole is described by a 
gauged sigma model. On the other hand, the work of \cite{BLO} suggests
that in order to put the M--theory KK--monopole in a massive background we
again have to use a gauged sigma model. In order to reproduce the
massive D-6-brane it is clear that in both cases the gauging must be
done in the isometry direction of the Taub--NUT space. However, we
can only gauge a given isometry direction once! The resolution of this
apparent puzzle is as follows. In a first stage, in order to describe the
massless M--theory KK--monopole we gauge the isometry direction using a 
{\it dependent}\footnote{In case the Wess-Zumino (WZ) term is ignored 
there is an 
alternative formulation of the gauged sigma model using an {\it independent}
worldvolume gauge field \cite{BJO}. 
However, in the presence of a WZ term the
two formulations are not equivalent and one must use the formulation
with the dependent gauge field.}
worldvolume gauge field $\hat A^{(1)}$ with gauged isometry
parameter $\hat\sigma^{(0)}$, see eqs.~(\ref{dep}) and (\ref{sigma}). 
We remind that
in order to put the M-wave, M-2-brane and M-5-brane in a massive
background we used a gauging with an {\it independent} gauge field
$\hat b$ \cite{BLO}. In the case of the M--theory KK--monopole, it
turns out that the dependent worldvolume field $\hat A^{(1)}$, when put in
a massive background, has exactly the same massive transformation
as the independent worldvolume field $\hat b$ we introduced for the
other branes. Therefore the dependent field $\hat A^{(1)}$ takes over the
role of the independent field $\hat b$.

Although in this sense the gauge field $\hat b$ is not needed 
to put the KK--monopole
in a massive background, it turns out that the worldvolume action
of the monopole already contains this field, even in a massless background. 
Whereas for the other branes it is an auxiliary field needed
to put the brane in a massive background, in the case of the
M--theory KK--monopole it plays the role of a propagating
Born--Infeld (BI) field.

Usually, a propagating worldvolume field describes the flux of another brane.
For instance, the BI 1--form in the D-brane worldvolume action
describes the flux of a fundamental string and the self--dual 2--form
in the M-5-brane worldvolume action describes the flux of an 
M-2-brane\footnote{The same worldvolume fields also play a role in the
construction of worldvolume solitons. For instance, the BI 1--form
corresponds to a 0--brane soliton and the self--dual 2--form is
used in constructing the self--dual string on the M-5-brane worldvolume.
Worldvolume solitons for the KK--monopole action have been considered
in \cite{george} recently.}. At this point a puzzle arises because
the 1--form field $\hat b$ appearing in the KK--monopole action
describes the
flux of a 1--brane and no such brane is known to exist in M--theory.
The resolution to this puzzle is that in fact the 1--form $\hat b$
describes the flux of a {\it wrapped} M-2-brane. One way to
see this is to observe that an M-2-brane can only intersect with an
M--theory KK--monopole over a 0-brane such that one of the
worldvolume directions of the M-2-brane coincides with the
isometry direction $z$ of the Taub-NUT space \cite{ber1}\footnote{
We use here a notation where a $\times (-)$ indicates a worldvolume (transverse)
direction. The $z$--direction corresponds to the isometry direction of the
Taub--NUT space and the first $\times$ in a row indicates the time
direction.}:

\begin{displaymath}
 (0|M2, \cal{KK})=
\mbox{ 
$\left\{ \begin{array}{c|cccccccccc}
         \x & \x & -  &  - & \x & - &  - & - & -  & -  & - \\
         \x & -& -& -&  z &\x & \x & \x& \x & \x & \x            
                                    \end{array} \right.$}   
\end{displaymath}

A compactification of this intersection 
along the $z$--direction gives the $(0|F1,D6)$
configuration. The reduction of $\hat b$ leads in this case
to the BI field of the 
D-6-brane describing the tension of the F1 fundamental string.

The organization of this letter is as follows. In Section 1 we present
the worldvolume action of the massive M--theory KK--monopole.
As a consistency check we verify in Section 2 that the direct reduction
of this action along $z$ gives the action of the massive D-6-brane.
Next, in Section 3 we perform a double dimensional reduction of the
M--theory KK--monopole and obtain the action
of the Type IIA KK--monopole. We perform this double dimensional
reduction only for the massless case leaving
the (more involved) massive case for a future publication.
Finally, in Section 4 we present our conclusions.


\section{The Action}

Before presenting the (bosonic) worldvolume action of the massive M--theory
KK--monopole, it is useful to first summarize the worldvolume 
and target space fields that are involved in the construction of this
action. We first discuss the worldvolume fields.
Since the monopole in 11 dimensions behaves like a 6--brane
and breaks half of the supersymmetry, it must correspond, after gauge fixing, 
to a 7--dimensional multiplet \cite{Hull}. The natural candidate is the
vector multiplet involving 3 scalars and 1 vector. As discussed in the
introduction, the embedding coordinates describe $11-7 = 4$ degrees of freedom
and one must eliminate a further scalar by gauging  an isometry of the
background.
Besides scalars and a vector, we also introduce a non-propagating
worldvolume 6-form that describes the tension of the
monopole\footnote{The same 6-form field occurs in the massive M-5-brane
action. In that case it leads to the spontaneous creation of a 
KK--monopole whenever an M-5-brane crosses an M-9-brane \cite{BLO}.}. 
The resulting worldvolume
field content of the KK--monopole is summarized in Table \ref{table1}.

\begin{table}[h]
\begin{center}
\begin{tabular}{||c|c|c|c||}
\hline\hline
Worldvolume     & Gauge         & Field         & $\sharp$ of   \\      
Field           & Parameter     & Strength      & d.o.f         \\
\hline\hline
${\hat X}^{{\hat \mu}}$ & & & $11 - 7 - (1) = 3$ \\
\hline
$ {\hat b}_{{\hat \imath}}$ & ${\hat \rho}^{(0)}$ &
 ${\hat {\cal F}}_{{\hat \imath}{\hat \jmath}}$& $7 - 2 = 5$\\
\hline
${\hat \omega}^{(6)}_{{\hat \imath}_1 \dots {\hat \imath}_6}$
& ${\hat \rho}^{(5)}$ 
& ${\hat {\cal K}}^{(7)}_{{\hat \imath}_1 \dots {\hat \imath}_7}$
& $-$ \\
\hline\hline
\end{tabular}
\end{center}  
\caption{\label{table1} {\bf Worldvolume Fields}. \footnotesize
In this table we give the 
worldvolume fields, together
with their gauge parameters,  field strengths
and number of degrees of freedom, that occur in the worldvolume
action of the massive M--theory KK--monopole. 
The worldvolume scalars ${\hat X}^{\hat \mu}$ are the embedding coordinates,
${\hat b}$ is a BI 1-form and
${\hat \omega}^{(6)}$ is a non propagating 6-form that describes the 
tension of the monopole. Due to the gauging the embedding scalars
describe 3 and not 4 degrees of freedom as indicated in the table.
}
\end{table}

We next discuss the target space fields. The target space background 
corresponds to massive 11-dimensional supergravity
\cite{BLO}, which has the same field content as the massless 11-dimensional
supergravity theory (see Table \ref{table2}).
The massive background has an isometry generated by a Killing vector
${\hat k}^{\hat\mu}$,
such that the Lie derivative of all target space fields and
gauge parameters with respect to ${\hat k}$ vanish\footnote{
We use the following notation for the hats.
Hats on target space fields and indices indicate they are
11-dimensional. Absence of hats indicates they are 10-dimensional.
Furthermore, 
we use hats for the worldvolume fields of branes in 11 dimensions
and no hats for the worldvolume fields of branes in 10 dimensions.
Finally, we use hats for worldvolume indices ${\hat \imath}$
if the  dimension  of the worldvolume is 6+1 and no hats
for a 5+1 dimensional worldvolume.}:

\begin{equation}
{\cal L}_{\hat k} {\hat C} = {\cal L}_{\hat k} 
{\hat \chi}= \dots =0 \, .
\end{equation}
\noindent Notice that in the massive case the gauge transformation rules
of the target space fields contain extra terms proportional to a
mass parameter $m$.

\begin{table}[h]
\begin{center}
\begin{tabular}{||c|c||}
\hline\hline
Target space & Gauge     \\
Field        & Parameter \\
\hline\hline
${\hat g}_{{\hat \mu}{\hat \nu}}$
& \\
\hline
${\hat C}_{{\hat \mu}{\hat \nu}{\hat \rho}}$ &
 ${\hat \chi}_{{\hat \mu}{\hat \nu}}$ \\
\hline
$(i_{\hat k} {\hat C})_{{\hat \mu}{\hat \nu}}$&
$(i_{\hat k} {\hat \chi})_{{\hat \mu}}$\\
\hline
${\hat {\tilde C}}_{{\hat \mu}_1 \dots {\hat \mu}_6}$
& ${\hat {\tilde \chi}}_{{\hat \mu}_1 \dots {\hat \mu}_5}$ \\
\hline
$\iktildeC_{{\hat \mu}_1 \dots {\hat \mu}_5}$
& $\iktildechi_{{\hat \mu}_1 \dots {\hat \mu}_4}$ \\
\hline
${\hat {\tilde N}}_{{\hat \mu}_1 \dots {\hat \mu}_8}$ 
& ${\hat \Sigma}_{{\hat \mu}_1 \dots {\hat \mu}_7}$ \\
\hline
$\iktildeN_{{\hat \mu}_1 \dots {\hat \mu}_7}$ & 
 $(i_{\hat k} {\hat \Sigma})_{{\hat \mu}_1 \dots {\hat \mu}_6}$\\
\hline\hline
\end{tabular}
\end{center}  
\caption{\label{table2} {\bf Target Space Fields}. \footnotesize
 This table shows the 11-dimensional target space
fields together with their gauge parameters. We also include
the contractions with the Killing vector ${\hat k}^{\hat \mu}$, denoted by
$(i_{\hat k} {\hat S})$ for a given field ${\hat S}$. 
The field ${\hat {\tilde C}}$ is the Poincar\'e dual of ${\hat C}$ and
${\hat {\tilde N}}$ is the Poincar\'e dual of the Killing vector, 
considered
as a 1-form ${\hat k}_{\hat \mu}$.
}
\end{table}

We proceed by presenting the (bosonic) action for the
massive M--theory KK--monopole:
\begin{equation}
\label{accion}
\begin{array}{rcl}
{\hat S}[{\hat X}^{\hat \mu}, {\hat b}_{\hat \imath}] &=&
-{\hat T}  \int d^7 \xi \,\,\,
{\hat k}^2 \sqrt{|{\rm det}(D_{{\hat \imath}} {\hat X}^{{\hat \mu}} 
D_{{\hat \jmath}} {\hat X}^{{\hat \nu}} 
{\hat g}_{{\hat \mu}{\hat \nu}}
+ \alfa {\hat k}^{-1} {\hat {\cal F}_{{\hat \imath} {\hat \jmath}}})|} \\
& & \\
& &
+ \,\, {1  \over 7!}
\alfa {\hat T} \int  d^7 \xi \,\, \epsilon^{{\hat \imath}_1 \dots {\hat \imath}_7}
\,\, {\hat {\cal K}}^{(7)}_{{\hat \imath}_1 \dots {\hat \imath}_7} \, . \\
\end{array}
\end{equation}

\noindent We use the following notation.
First of all, $\hat {k}^2 = -\hat {k}^{\hat \mu}\hat {k}^{\hat \nu}
\hat{g}_{\hat\mu\hat\nu}$. The field--strength $\hat {\cal F}$
of the BI 1--form $\hat b$ is defined by\footnote{
We omit the wordvolume indices. It is understood that the expression is
completely antisymmetrized in these indices. We use this notation
throughout the paper.}

\begin{equation}
\label{efe}
\begin{array}{rcl}
{\hat {\cal F}} &=& 2 \partial {\hat b} + {1 \over 2\pi\alpha^{\prime}}
\partial {\hat X}^{{\hat \mu}} \partial {\hat X}^{{\hat \nu}}
\ikC_{{\hat \mu}{\hat \nu}} \, , \\
\end{array}
\end{equation}

\noindent where $(i_{\hat k} {\hat C})$ is the contraction 
of the field ${\hat C}$ with ${\hat k}$. The covariant derivative is given by
\begin{equation}
D_{\hat \imath}{\hat X}^{\hat \mu} = \partial_{\hat \imath}
{\hat X}^{\hat \mu} + {\hat A}_{\hat \imath}^{(1)} 
{\hat k}^{\hat \mu}\, ,
\end{equation}
where the  gauge field ${\hat A}^{(1)}$  is a dependent
field given by
\begin{equation}
\label{campo}
{\hat A}_{\hat \imath}^{(1)}
  = {\hat k}^{-2} \partial_{\hat \imath} 
{\hat X}^{{\hat \mu}} {\hat k}_{\hat \mu} \, .
\label{dep}
\end{equation}

\noindent Finally, ${\hat {\cal K}}^{(7)}$ is the field strength of the
non propagating worldvolume 6-form ${\hat \omega}^{(6)}$:
\begin{equation}
\begin{array}{rcl}
{\hat {\cal K}}^{(7)} &=&
7 \left\{ \partial {\hat \omega}^{(6)} - {1 \over 7 \alfa}
({i}_{\hat k} {\hat {\tilde N}}) 
+3 ({i}_{\hat k}{\hat {\tilde C}}){\hat {\cal F}} \right.
\\ & & \\ 
& &
- {5 \over 2\pi\alpha^{\prime}}D  {\hat X}^{{\hat \mu}}
D  {\hat X}^{{\hat \nu}}D  {\hat X}^{{\hat \rho}}
{\hat C}_{{\hat \mu} {\hat \nu} {\hat \rho}} \ikC  \ikC \\
& &\\
& &
-30 D {\hat X}^{{\hat \mu}}D {\hat X}^{{\hat \nu}}
D {\hat X}^{{\hat \rho}}
{\hat C}_{{\hat \mu} {\hat \nu} {\hat \rho}} \ikC  \partial {\hat b}
\\ & & \\ 
& &
-60 \alfa D {\hat X}^{{\hat \mu}} D {\hat X}^{{\hat \nu}}
D {\hat X}^{{\hat \rho}} {\hat C}_{{\hat \mu} {\hat \nu}
{\hat \rho}}
\partial {\hat b} \partial {\hat b}  \\
& &\\
& &
\left. -120\alfa^2 {\hat A}^{(1)} \partial {\hat b} \partial {\hat b} 
\partial {\hat b}
-15 m \alfa^3 {\hat b} \partial {\hat b} \partial {\hat b} 
\partial {\hat b}
\right\} \, . \\
\end{array}
\end{equation}
\noindent Notice that we write explicitly the pullbacks only when
covariant derivatives are used.

The full action is invariant under worldvolume gauge transformations
(see Table 1), target space gauge transformations 
(see Table 2) and local isometry transformations (with parameter 
$\hat\sigma^{(0)}$) as given below.
The gauge transformations of the worldvolume fields 
are given by:
\begin{equation}
\begin{array}{rcl}
\delta {\hat X}^{\hat \mu} &=& -
{\hat \sigma}^{(0)} {\hat k}^{\hat \mu}\, ,\\
& & \\
\delta {\hat b} &=& 
 \partial {\hat \rho}^{(0)} - {1 \over 2\pi\alpha^{\prime}}
\ikchi \, ,\\
& &\\
\delta {\hat \omega}^{(6)}
 &=& 6 \partial{\hat \rho}^{(5)} + {1 \over 2\pi\alpha^{\prime}}
 (i_{\hat k} {\hat \Sigma})
- 30{\hat b} \partial
 ({\hat \imath}_{\hat k} {\hat {\tilde \chi}})
\\ & & \\ & &
-180 \alfa \partial
 {\hat \chi} {\hat b}
 \partial {\hat b}
- 120 \alfa^2 \partial {\hat \sigma}^{(0)} 
 {\hat b} \partial
 {\hat b} \partial
 {\hat b}  
\\ & & \\ & &
- 45m \alfa^2 \ikchi
 {\hat b} \partial
 {\hat b} \partial
 {\hat b} 
- 15 m \alfa^3 \partial {\hat \rho}^{(0)}
 {\hat b}
 \partial
 {\hat b}
 \partial
 {\hat b} \, .\\
\end{array}
\end{equation}
\noindent The dependent gauge field ${\hat A}^{(1)}$ transforms as
\begin{equation}
\delta {\hat A}^{(1)} = \partial {\hat \sigma}^{(0)} + {m \over 2}
( i_{\hat k} {\hat \chi}) \, .
\label{sigma}
\end{equation}
Note that $\hat\omega^{(6)}$ transforms both under
$\hat\sigma^{(0)}$ and $\hat\rho^{(0)}$. Only if we identify the two
parameters

\begin{equation}
{\hat \sigma}^{(0)} = - {m \over 2} \alfa {\hat \rho}^{(0)}\, ,
\label{ident}
\end{equation}
the transformation rule of $\hat\omega^{(6)}$ coincides with the one given in 
\cite{BLO} where it occurred in the coupling to a massive M-5-brane.
We deduce that the ${\hat \omega}^{(6)}$ describing the monopole
tension actually has more symmetries 
than when it is coupled to a massive M-5-brane. The
reason for this is that in the case of the M-5-brane
the gauge transformation of $\hat b$ is identified with the gauged
isometry parameter.

\noindent Finally, the gauge transformations of the target space fields are
given by:
\begin{displaymath}
\begin{array}{rcl}
\label{transf}
\delta {\hat g}_{{\hat \mu} {\hat \nu}} &=& 
 - m \ikchi_{({\hat \mu}} (i_{\hat k} {\hat g})_{{\hat \nu})}
- {\hat \sigma}^{(0)}
 {\hat k}^{\hat \lambda} 
\partial_{\hat \lambda} {\hat g}_{{\hat \mu}{\hat \nu}}
\, ,\\
& & \\
\delta {\hat C}_{{\hat \mu}{\hat \nu} {\hat \rho}}
 &=& 3 \partial_{[ {\hat \mu}} {\hat \chi}_{ {\hat \nu} {\hat \rho}]}
 - {3 \over 2}m \ikchi_{[{\hat \mu}} \ikC_{{\hat \nu}{\hat \rho}]} 
- {\hat \sigma}^{(0)}
{\hat k}^{\hat \lambda} \partial_{\hat \lambda} 
{\hat C}_{{\hat \mu}{\hat \nu}{\hat \rho}} 
\, ,\\
& & \\
\delta \ikC_{{\hat \mu}{\hat \nu}}  &=& 2 \partial_{[{\hat \mu}}
 \ikchi_{{\hat \nu}]} - {\hat \sigma}^{(0)}
{\hat k}^{\hat \lambda} \partial_{\hat \lambda} \ikC_{{\hat \mu}{\hat \nu}}
 \, ,\\
& & \\
\delta \iktildeC_{{\hat \mu}_1 \dots {\hat \mu}_5}
  &=&5 \partial_{[{\hat \mu}_1} 
\iktildechi_{{\hat \mu}_2 \dots {\hat \mu}_5]}
+ 15 \partial_{[{\hat \mu}_1} {\hat \chi}_{{\hat \mu_2}{\hat \mu}_3} 
\ikC_{{\hat \mu}_4 {\hat \mu}_5]}\\
& & \\
&&- 10{\hat C}_{[{\hat \mu}_1 {\hat \mu}_2 {\hat \mu}_3} \partial_{{\hat \mu}_4}
 \ikchi_{{\hat \mu}_5]} 
- {\hat \sigma}^{(0)} {\hat k}^{\hat \lambda}
\partial_{\hat \lambda} \iktildeC_{{\hat \mu}_1 \dots {\hat \mu}_5}  \, , \\
& & \\
\delta \iktildeN_{{\hat \mu}_1 \dots {\hat \mu}_7}  &=&
7 \left\{ \partial_{[{\hat \mu}_1} 
(i_{\hat k} {\hat  \Sigma})_{{\hat \mu}_2 \dots {\hat \mu}_7]} 
+15 \partial_{[{\hat \mu}_1} \iktildechi_{{\hat \mu}_2 \dots
{\hat \mu}_5} \ikC_{{\hat \mu}_6 {\hat \mu}_7]} \right.\\
& &\\
& & \left.+30 \partial_{[{\hat \mu}_1}
 {\hat \chi}_{{\hat \mu}_2 {\hat \mu}_3}
 \ikC_{{\hat \mu}_4 {\hat \mu}_5} 
 \ikC_{{\hat \mu}_6 {\hat \mu}_7]}\right.\\
& &\\
& & \left.
 -20 {\hat C}_{[{\hat \mu}_1 {\hat \mu}_2 {\hat \mu}_3}
\ikC_{{\hat \mu}_4 {\hat \mu}_5} \partial_{{\hat \mu}_6}
\ikchi_{{\hat \mu}_7]} \right\}
-{\hat \sigma}^{(0)} {\hat k}^{\hat \lambda} 
\partial_{\hat \lambda} 
\iktildeN_{{\hat \mu}_1 \dots {\hat \mu}_7}
 \, .\\
\end{array}
\end{displaymath}
\noindent


\section{Direct Dimensional Reduction}


As a consistency check on the monopole action presented in the
previous Section we perform a direct dimensional reduction along the 
direction associated
to the isometry of the Taub-NUT space. 
It is convenient to use adapted coordinates  
${\hat k}^{\hat \mu}= \delta^{{\hat \mu} z}$.
The embedding scalars then split 
\begin{equation}
{\hat X}^\mu = X^\mu \, , \qquad {\hat X}^z = Z \, ,
\end{equation}
\noindent and the background fields reduce as follows\footnote{We use the 
notation and conventions of \cite{BLO}.}:
\begin{displaymath}
\label{dimred}
\begin{array}{rcl}
{\hat g}_{\mu \nu} &=& e^{- {2 \over 3} \phi} g_{\mu \nu}
- e^{{4 \over 3} \phi} C^{(1)}_\mu C^{(1)}_\nu \, ,\\
& &\\
{\hat g}_{\mu z} &=& - e^{{4 \over 3}\phi} C^{(1)}_\mu \, ,\\
& & \\
\iktildeC_{\mu_1 \dots \mu_5} &=& C^{(5)}_{\mu_1 \dots \mu_5}
 - 5 C^{(3)}_{[\mu_1 \mu_2 \mu_3}B_{\mu_4 \mu_5]} \, , \\
& & \\
\iktildeN_{\mu_1 \dots \mu_7} &=& C^{(7)}_{\mu_1 \dots \mu_7}
 - 5 \cdot 7 C^{(3)}_{[\mu_1 \mu_2 \mu_3} 
B_{\mu_4 \mu_5}B_{\mu_6 \mu_7]} \, .\\
\end{array}
\begin{array}{rcl}
{\hat g}_{zz} &=& -e^{{4 \over 3} \phi} \, ,\\
& & \\
{\hat C}_{\mu \nu \rho} &=& C^{(3)}_{\mu \nu \rho} \, , \\
& & \\
\ikC_{\mu \nu} &=& B_{\mu \nu} \, ,\\
\end{array}
\end{displaymath}
Similarly, the reduction rules of the gauge parameters are given by
\begin{equation}
\begin{array}{rcl}
\iktildechi_{\mu_1 \dots \mu_4} &=& \Lambda^{(4)}_{\mu_1 \dots \mu_4} \, ,\\
& & \\
{\hat \chi}_{\mu \nu} &=& \Lambda^{(2)}_{\mu \nu} \, ,\\
\end{array}
\hspace{.2cm}
\begin{array}{rcl}
\ikchi_{\mu} &=& \Lambda_{\mu} \, ,\\
& & \\
(i_{\hat k}{\hat \Sigma})_{\mu_1 \dots \mu_6} &=& 
\Lambda^{(6)}_{\mu_1 \dots \mu_6}\, .\\
\end{array}
\end{equation}
Besides these gauge parameters, we include as well the gauge
parameter $\Lambda^{(0)}$ associated to 
the KK vector $C^{(1)}$.

Concerning the worldvolume fields, the BI  field ${\hat b}$ reduces as
\begin{equation}
{\hat b}_{\hat \imath}= b_{\hat \imath} \, .
\end{equation}

\noindent Furthermore, we have

\begin{equation}
{\hat A}_{\hat \imath}^{(1)} = - C_{\hat \imath}^{(1)} - 
\partial_{\hat \imath} Z \, ,\qquad
D_{\hat \imath} {\hat X}^{z} = - C_{\hat \imath}^{(1)} \, ,\qquad
D_{\hat \imath} {\hat X}^{\mu} =
\partial_{\hat \imath} X^\mu \, . 
\end{equation}
Substituting the above reduction rules into the kinetic term of the
monopole action one gets
the usual kinetic term for $D$-branes
\begin{equation}
\begin{array}{rcl}
{\hat k}^2
\sqrt{|{\rm det}(D_{{\hat \imath}} {\hat X}^{\hat \mu} D_{{\hat \jmath}} 
{\hat X}^{\hat \nu}  {\hat g}_{{\hat \mu}{\hat \nu}} 
+ \alfa {\hat k}^{-1} {\hat {\cal F}}_{{\hat \imath}{\hat \jmath}})|} &=&
\\ & & \\ & &
\hspace{-4.5cm}
e^{-  \phi} \sqrt{ |{\rm det}( \partial_{{\hat \imath}}{X}^\mu
 \partial_{{\hat \jmath}} {X}^\nu g_{\mu \nu} + \alfa 
{\cal F}_{{\hat \imath}{\hat \jmath}})|} \, ,\\
\end{array}
\end{equation}
\noindent where 
\begin{equation}
{\cal F}= 2 \partial b + {1 \over 2\pi\alpha^{\prime}}B 
\end{equation}
is the BI curvature.

We next discuss the reduction of $\hat\omega^{(6)}$. Since the KK--monopole 
reduces to the D-6-brane it is expected that $\hat\omega^{(6)}$ will
reduce to the field
$c^{(6)}$ describing the tension of the D-6-brane.
The gauge transformations of $c^{(6)}$ are  given by \cite{BLO}:
\begin{equation}
\begin{array}{rcl}
\delta c^{(6)} &=& 6 \partial \kappa^{(5)} - {1 \over 2\pi\alpha^{\prime}}  
\Lambda^{(6)} + 30 \Lambda^{(4)}\partial b  
- 180 \alfa \Lambda^{(2)} \partial b\partial b 
\\
& & \\
& & + 120 \alfa^2 \Lambda^{(0)} \partial b \partial b \partial b 
-15 m \alfa^3 \rho^{(0)} \partial b \partial b \partial b 
\\
& & \\
& & + 45m \alfa^2 \Lambda b \partial b \partial b\, .
\end{array}
\end{equation}
We find the following reduction rule for ${\hat \omega}^{(6)}$\footnote{
This reduction rule coincides with the one given in \cite{BLO} provided
we make the identification (\ref{ident}) and rename $Z = (2\pi
\alpha^\prime) c^{(0)}$.}

\begin{equation}
{\hat \omega}^{(6)} = - c^{(6)} + 120 \alfa^2 \partial Z b \partial b 
\partial b \, .
\label{omega(g)-direct-red}
\end{equation}
\noindent Making use of the above reduction rules we find that,
after direct reduction,
the coordinate $Z$ disappears from the action and we obtain the
action of a massive $D6$-brane :

\begin{equation}
\begin{array}{rcl}
S[X^\mu, b_{\hat \imath}] &=&
- {\hat T} \int d^7 \xi \,\,
e^{-\phi} \sqrt{|{\rm det}(g+ \alfa {\cal F})|}\\
& & \\
& &
- \,\, {1 \over 7!}\alfa {\hat T} \int d^7\xi   
\,\,  \epsilon^{{\hat \imath}_1 \dots {\hat \imath}_7}
{\cal G}^{(7)}_{{\hat \imath}_1 \dots {\hat \imath}_7} \, .\\
\end{array}
\end{equation}
\noindent Here ${\cal G}^{(7)}$ is the curvature
of the worldvolume field $c^{(6)}$:

\begin{equation}
\begin{array}{rcl}
{\cal G}^{(7)} &=& 7 \,\,
\left\{ \partial c^{(6)} + {1 \over 7 \alfa} C^{(7)} 
  - 3 C^{(5)} {\cal F} \right. 
 + 15\alfa C^{(3)} {\cal F}{\cal F}
\\ 
& &\\
& &
\left. - 15 \alfa^2 C^{(1)} {\cal F}{\cal F}{\cal F} 
+ 15 m \alfa^3 b \partial b \partial b \partial b \, \right\}\, .
\\
\end{array}
\end{equation}

\noindent The gauge transformations of the worldvolume and target space 
fields coupled to the $D$-6-brane can be found,
for instance, in \cite{BLO}.


\section{Double Dimensional Reduction}


In this Section
we perform the double dimensional reduction of the M-theory KK-monopole
given in Section 1. As a result we will obtain
the action of the Type IIA KK--monopole. We consider 
only the massless case\footnote{For a discussion of the massive case,
see the Conclusions.}, i.e.~our starting point is 
the action (\ref{accion}) with $m=0$.
In order to perform the double dimensional reduction, we must introduce
an extra isometry for the background, 
generated by a Killing vector ${\hat h}$, and then
wrap one direction of the monopole, ${\hat \xi}^6$, around
this new compact direction:
\begin{equation}
\partial_6{\hat X}^{\hat \mu} = {\hat h}^{\hat \mu} \, .
\end{equation}
We have now two different isometries, one is given by 
${\hat h}$, which is in a direction
tangent to the worldvolume, and the other one is given by ${\hat k}$,
in a direction transverse to the worldvolume.
For the action (\ref{accion}) to be invariant under both isometries we must
require that ${\cal L}_{\hat h} {\hat k} =0$. 
Moreover, this implies that we can find a coordinate 
system adapted to both isometries\footnote{Notice, that the usage of the
adapted coordinates does not imply that 
${\hat k}^{\hat \mu} {\hat h}^{\hat \nu} {\hat g}_{{\hat \mu}{\hat \nu}} = 
0$.}: ${\hat h}^{\hat \mu} = \delta^{{\hat \mu} y}$, 
${\hat k}^{\hat \mu} = \delta^{{\hat \mu} z}$. 
Splitting the embedding coordinates as
${\hat X}^{\hat \mu} = (X^\mu, Y)$
we find the gauge-fixing condition:
\begin{equation}
Y= {\hat \xi}^6\, .
\end{equation}
In this double dimensional reduction the 
Killing vector ${\hat k}$ becomes, after reduction,
the Killing vector associated to the isometry of the
Taub-NUT space of the IIA KK-monopole:
\begin{equation}
{\hat k}^y = 0 \, ,\qquad {\hat k}^\mu = k^\mu \, .
\end{equation}

We use the following reduction rules for the background fields:
\begin{displaymath}
\begin{array}{rcl}
{\hat g}_{\mu \nu}&=& e^{- {2 \over 3} \phi} g_{\mu \nu} -
e^{ {4 \over 3} \phi} C^{(1)}_\mu C^{(1)}_\nu \, ,\\ 
& & \\
{\hat g}_{\mu y} &=& - e^{ {4 \over 3} \phi} C^{(1)}_\mu \, ,\\ 
& & \\
{\hat g}_{yy} &=& - e^{ {4 \over 3} \phi} \, ,\\ 
& & \\
\iktildeN_{\mu_1 \dots \mu_6 y} &=& -(i_k {\tilde N})_{\mu_1 \dots \mu_6}
 \, ,\\
& & \\
{\hat {\tilde C}}_{\mu_1 \dots \mu_5 y} &=& 
C^{(5)}_{\mu_1 \dots \mu_5} - 5C^{(3)}_{[\mu_1 \mu_2 \mu_3}
B_{\mu_4 \mu_5]} \, ,\\
& & \\
\iktildeC_{\mu_1 \dots \mu_4 y} &=& -(i_k C^{(5)})_{\mu_1 \dots \mu_4}
 + 3 (i_k C^{(3)})_{[\mu_1 \mu_2} B_{\mu_3 \mu_4]} +
2 C^{(3)}_{[\mu_1 \mu_2 \mu_3} (i_k B)_{\mu_4]} \, ,\\
\end{array}
\hspace{-4.5cm}
\begin{array}{rcl}
{\hat C}_{\mu \nu \rho} &=& C^{(3)}_{\mu \nu \rho} \, ,\\ 
& & \\
{\hat C}_{\mu \nu y} &=& B_{\mu \nu} \, , \\ 
& & \\
\ikC_{\mu y} &=& - (i_k B)_\mu \, ,\\ 
& & \\
{\hat {\tilde C}}_{\mu_1 \dots \mu_6} &=& - {\tilde B}_{\mu_1 \dots \mu_6}
 \, ,\\
\end{array}
\end{displaymath}
where ${\hat {\tilde N}}$ reduces to ${\tilde N}$, the Poincar{\'e} dual
of the Killing vector, considered as a 1-form $k_\mu$.
\noindent Similarly, the reduction rules of the gauge 
parameters are given by
\begin{equation}
\begin{array}{rcl}
{\hat {\tilde \chi}}_{\mu_1 \dots \mu_5} &=& 
- {\tilde \Lambda}_{\mu_1 \dots \mu_5} \, ,\\ 
& & \\
{\hat {\tilde \chi}}_{\mu_1 \dots \mu_4 y} &=& \Lambda^{(4)}_{\mu_1
\dots \mu_4} \, ,\\
& & \\
\iktildechi_{\mu_1 \dots \mu_3 y} &=& -(i_k \Lambda^{(4)})_{\mu_1 \dots \mu_3}
 \, ,\\ 
& & \\
(i_{\hat k}{\hat {\tilde \Sigma}})_{\mu_1 \dots \mu_5 y} &=& 
- (i_k \Sigma^{(6)})_{\mu_1 \dots \mu_5} \, .\\
\end{array}
\hspace{1.2cm}
\begin{array}{rcl}
{\hat \chi}_{\mu y} &=& \Lambda_\mu \, ,\\ 
& & \\
\ikchi_{y} &=& (i_k \Lambda) \, , \\ 
& & \\
\ikchi_{\mu} &=& (i_k \Lambda^{(2)})_{\mu} \, . \\ 
& & \\
\end{array}
\end{equation}

We now turn to the reduction of the gauge field
${\hat A}^{(1)}$. Using the reduction rules given above
one obtains:
\begin{equation}
\begin{array}{rcl}
{\hat A}^{(1)}_i &=& \left( 1 + e^{2 \phi} k^{-2} (i_k C^{(1)})^2 \right)^{-1}
\left( A^{(1)}_i - e^{2 \phi}k^{-2}
C^{(1)}_i (i_k C^{(1)}) \right)
\, , \\ & & \\
{\hat A}^{(1)}_6 &=& -
\left( 1 + e^{2 \phi} k^{-2} (i_k C^{(1)})^2 \right)^{-1}
 e^{2 \phi} k^{-2} (i_k C^{(1)}) \, ,\\
\end{array}
\end{equation}
\noindent where $k^2= - k^\mu k^\nu g_{\mu \nu}$ and the field $A^{(1)}$
has been defined as the dependent gauge
field associated to the gauged isometry of the Type IIA KK--monopole:

\begin{equation}
A^{(1)}_i = k^{-2} \partial_i X^\mu k_\mu\, .
\end{equation}
The covariant derivative is defined by
\begin{equation}
DX^\mu = \partial X^\mu + A^{(1)} k^\mu \, .
\end{equation}

Concerning the worldvolume fields, the 1-form ${\hat b}$ 
gives rise to a scalar and a 1-form, and the 6-form 
${\hat \omega}^{(6)}$ reduces to a 5-form $\omega^{(5)}$ describing
the tension of the IIA KK--monopole:
\begin{equation}
{\hat b}_i = \omega^{(1)}_i \, ,\qquad
{\hat b}_6 = \omega^{(0)} \, ,\qquad
{\hat \omega}^{(6)}_{i_1 \dots i_5 6} = 
\omega^{(5)}_{ i_1 \dots i_5} \, .
\end{equation}

Using the above reduction rules in the action of the
massless M--theory KK--monopole, we find the following action of 
a massless Type IIA KK--monopole:

\begin{equation}
\label{m0accion}
\begin{array}{rcl}
S[X^\mu, \omega^{(0)}, \omega^{(1)}] &=&
-T \int d^6 \xi \,\,
e^{-2\phi} k^2 \sqrt{1 + e^{2 \phi}k^{-2} (i_k C^{(1)})^2} \times \\
& & \\
& &\hspace{-3.5cm}
\times\sqrt{|{\rm det}(D_iX^\mu D_jX^\nu g_{\mu \nu}
- \alfa^2 k^{-2} {\cal K}^{(1)}_i {\cal K}^{(1)}_j +
 { \alfa  k^{-1} e^\phi \over \sqrt{ 1 + e^{2 \phi}k^{-2} 
(i_k C^{(1)})^2}}
{\cal K}^{(2)}_{ij})|}\\
& & \\
& &\hspace{-3.5cm}
+ \,\, {1 \over 6!}\alfa T \int d^6\xi
\,\,  \epsilon^{i_1 \dots i_6}
{\cal K}^{(6)}_{i_1 \dots i_6} \, .\\
\end{array}
\end{equation}
\noindent Here $T$ is the tension of the Type IIA KK--monopole,
which is related to the tension of the M--theory KK--monopole by:
\begin{equation}
T = {\hat T} \int_{S^1} d {\hat \xi}^6 \, .
\end{equation}
\noindent The 1-form  ${\cal K}^{(1)}$ is 
the field strength of the scalar $\omega^{(0)}$
\begin{equation}
{\cal K}^{(1)} = \partial \omega^{(0)} - {1 \over 2\pi\alpha^{\prime}}
(i_k B) \, ,
\end{equation}
\noindent and the field strength ${\cal K}^{(2)}$ of the 
1-form  $\omega^{(1)}$ is given by
\begin{equation}
{\cal K}^{(2)}= 2 \partial \omega^{(1)} + {1 \over 2\pi\alpha^{\prime}}
(i_k C^{(3)})
- 2 {\cal K}^{(1)} \left(C^{(1)}+ (i_k C^{(1)}) A^{(1)}\right) \, .
\end{equation}
\noindent Finally, ${\cal K}^{(6)}$ 
is the field strength of the non propagating
worldvolume 5-form $\omega^{(5)}$:
\begin{equation}
\begin{array}{rcl}
{\cal K}^{(6)} &=& \left\{ 6 \partial \omega^{(5)} 
+\frac{1}{2\pi\alpha^\prime}(i_k {\tilde N})-
30 (i_k C^{(5)})\partial\omega^{(1)}-\frac{15}{2\pi\alpha^\prime}
(i_k C^{(5)})(i_k C^{(3)})\right.
\\ & & \\ & &
-6(i_k {\tilde B}){\cal K}^{(1)}
 -120 \alfa DX^{{\mu}} DX^{{\nu}} DX^{{\rho}} 
C^{(3)}_{{\mu} {\nu} {\rho}} 
\, {\cal K}^{(1)} \partial \omega^{(1)}
\\ & & \\ & & 
+{30 \over 2\pi\alpha^{\prime}} DX^{\mu}DX^{\nu} B_{\mu \nu}
 (i_k C^{(3)}) (i_k C^{(3)})
\\ & & \\ & & 
+ {50 \over 2\pi\alpha^{\prime}} DX^{\mu} DX^{\nu} DX^{\rho} 
C^{(3)}_{\mu \nu \rho} 
(i_k B) (i_k C^{(3)})
\\ & & \\ & & 
-30DX^{\mu} DX^{\nu} DX^{\rho} C^{(3)}_{\mu \nu \rho} 
( i_k C^{(3)}) \partial \omega^{(0)}
\\ & & \\ & & 
-180 \alfa DX^{\mu} DX^{{\nu}} B_{{\mu} {\nu}} 
\partial \omega^{(1)}\partial \omega^{(1)}
\\ & & \\ & & 
-360 \alfa^2 A^{(1)} \partial \omega^{(1)} \partial \omega^{(1)} 
\partial \omega^{(0)}
\\ & & \\ & & 
\left. + 15 \alfa^2 { e^{2 \phi} k^{-2} (i_k C^{(1)}) \over 
1 + e^{2 \phi} k^{-2} (i_k C^{(1)})^2}
{\cal K}^{(2)} {\cal K}^{(2)} {\cal K}^{(2)} \right\} \, ,\\
\end{array}
\end{equation}
\noindent where again, we are writing 
the pullbacks explicitly only when covariant
derivatives are used.

The action is invariant under (massless) gauge
transformations and the gauged isometry. 
The transformations for the worldvolume fields are:
\begin{equation}
\begin{array}{rcl}
\delta X^\mu &=& -\sigma^{(0)} k^\mu \, \\
& & \\
\delta \omega^{(0)} &=& {1 \over 2\pi\alpha^{\prime}} (i_k \Lambda) \, ,\\
& & \\
\delta \omega^{(1)} &=& \partial \rho^{(0)} - {1 \over 2\pi\alpha^{\prime}}
(i_k \Lambda^{(2)}) + \partial \Lambda^{(0)} \omega^{(0)} \, , \\
& & \\
\delta \omega^{(5)} &=& 5 \partial \rho^{(4)} -
{1 \over 2\pi\alpha^{\prime}} (i_k \Sigma^{(6)}) 
- 5 \partial(i_k {\tilde \Lambda}) \omega^{(0)}
+20 \, \omega^{(1)} \partial (i_k \Lambda^{(4)}) \\
& & \\
& &-30 \alfa \partial \Lambda^{(2)} (\omega^{(1)} \partial \omega^{(0)}
 + \partial \omega^{(1)} \omega^{(0)})
 \\
& & \\
& & + 60 (2\pi\alpha^\prime)\omega^{(1)} \partial 
\omega^{(1)} \partial \Lambda +
60 \alfa^2 \sigma^{(0)} \partial \omega^{(1)} \partial \omega^{(1)}
 \partial \omega^{(0)} \, ,
\\
\end{array}
\end{equation}

\noindent whereas that of
the target space field $(i_k {\tilde N})$ is given by:

\begin{equation}
\begin{array}{rcl}
\delta (i_k {\tilde N})
 &=& 6 \partial (i_k \Sigma^{(6)})
+60 (i_k C^{(3)}) \partial
 (i_k \Lambda^{(4)})
- 30 \partial (i_k {\tilde \Lambda})(i_k B)
\\ & & \\ & &
- 60 (i_k C^{(3)})(i_k C^{(3)})\partial \Lambda  
+ 120 \partial \Lambda^{(2)}(i_k C^{(3)}) (i_k B) \\
& & \\
& &
 +60 (i_k C^{(3)}) \partial
( i_k \Lambda^{(2)}) B 
- 40 C^{(3)} \partial
 (i_k \Lambda^{(2)}) (i_k B) \\
& & \\
& &
- 20 C^{(3)} (i_k C^{(3)})
\partial (i_k \Lambda) 
- \sigma^{(0)} k^\lambda \partial_\lambda (i_k {\tilde N})
\, .\\
\end{array}
\end{equation}
\noindent The transformations of the other target space fields can be
found in \cite{BLO}.

Notice that the truncation of the action (\ref{m0accion}) by
setting the $RR$ fields and 
${\omega}^{(1)}$ to zero leads to the action 
of the Heterotic KK-monopole given in \cite{Bert}.


\section{Conclusions}

In this paper we have presented the worldvolume
action of the massive M--theory
KK--monopole. We find that the action is given by a gauged sigma-model. 
This agrees with the results of \cite{BLO} where 
the effective actions of massive branes where generally shown 
to be described by gauged sigma-models.
 
A new feature that arises is that
the M--theory KK--monopole action is already given by
a gauged sigma-model in the massless case.
This gauging is needed to
effectively eliminate the Taub-NUT isometry coordinate
\cite{BJO}. We find that the dependent gauge field
that is needed for the gauging in the massless case
transforms under the massive transformations such that the massless 
gauged action
can be made invariant under the massive transformations without the
need to introduce an independent auxiliary gauge field. Besides a
dependent gauge field, the
monopole action contains an independent BI gauge field, which
already occurs in the massless case. It turns out that in the
massive case extra terms 
proportional to the mass parameter $m$
need to be added to the WZ term. 
As a check on the action we have presented in this work, 
we have performed a direct dimensional reduction along the
Taub--NUT isometry direction and have obtained the massive 
D-6-brane.

We have as well performed a double dimensional reduction and obtained the
action of the Type IIA KK--monopole. We have performed this double
dimensional reduction only for the massless case. In a forthcoming
publication we will perform the double dimensional reduction
in the massive case and present the massive Type IIA KK--monopole 
\cite{tutti}. A new feature that arises in the double
dimensional reduction of the massive monopole is that 
we first need to construct a massive
monopole in which the isometry associated to the mass
is realized along a coordinate {\it different} from the Taub-NUT 
isometry direction. We thus end up with an M--theory
KK--monopole with two gauged isometries. One isometry
is the $z$--direction in the Taub--NUT space. This isometry is
already gauged in the massless case. The other isometry is only
gauged when the monopole is put in a massive background. 

Finally, our results for the Type IIA monopole also lead, via T--duality,
to the worldvolume action of a Type IIB NS5--brane, which coincides
with that of a D-5-brane in a SL(2,Z)-transformed background,
see \cite{tutti}.
In order to work out this duality transformation it is necessary
to first derive the set of T-duality rules which apply to the different
world-volume fields that need to be introduced.


\section*{Acknowledgments}

We would like to thank A.~Ach\'ucarro, 
J.M.~Figueroa-O'Farrill, B.~Janssen and 
J.P.~van der Schaar for useful 
discussions. E.E.~ thanks QMW college for its hospitality.
The work of E.B.~is supported by the European Commission TMR program
ERBFMRX-CT96-0045, in which E.B.~is associated to the University of
Utrecht. The work of E.E.~ is part of the research program of the
"Stich\-ting FOM".
The work of Y.L.~is supported by a TMR fellowship from the
European Commission.


\end{document}